\documentstyle[aps,preprint]{revtex}
\begin{document}
\draft
\title{ Chern-Simons Field Theory of Two-Dimensional
Electrons in the Lowest Landau Level}
\author{
Lizeng Zhang }
\address{
Department of Physics and Astronomy \\
The University of Tennessee, Knoxville, TN 37996\\
and Solid State Division, Oak Ridge National Laboratory\\
Oak Ridge, Tennessee 37831\\
}

\maketitle

\begin{abstract}
We propose a fermion Chern-Simons field theory describing
two-dimensional electrons in the lowest Landau level. This
theory is constructed with a complete
set of states, and the lowest Landau level constraint is
enforced through a $\delta$-functional described by an
auxiliary field $\lambda$. Unlike the field theory constructed
directly with the states in the lowest Landau level, this
theory allows one utilizing the physical picture of
``composite fermion" to study the fractional quantum Hall
states by mapping them onto certain integer quantum Hall
states; but unlike it in the unconstrained theory, such
a mapping is sensible only when interactions between
electrons are present. An ``effective mass", which
characterizes the scale of low energy excitations in the
fractional quantum Hall systems, emerges naturally from
our theory. We study a Gaussian effective theory
and interpret physically the dressed stationary point
equation for $\lambda$ as an equation for the ``mass
renormalization" of composite fermions.
\end{abstract}

\section{Introduction}

The fermion Chern-Simons field theoretical approach
\cite{LF,HLR}, motivated by the composite fermions theory
\cite{Jain} in which the fractional quantum Hall effect
(FQHE) \cite{book} is viewed as the integer QHE of certain
bound objects of electrons and vortices, i.e, the ``composite
fermions", has shown to be a useful method for investigating
the physical properties of the FQHE. In such a field
theoretical approach, one, roughly speaking, ``attaches" even
number of flux quanta to the two-dimensional (2D) electrons
through introducing the so called Chern-Simons gauge field.
In the mean-field approximation where the statistical gauge
fluxes are delocalized from the electrons and uniformly spread out
in the 2D plane, this mean gauge field partially cancels the
external magnetic field such that at certain filling fractions
the mean-field state is descibed by integer number of filled
effective Landau level (LL), which signals the occurrence
of the FQHE.  Qualitatively correct physics of the FQHE may
be recovered by incorporating the effect of fluctuations of
the gauge field.  This approach has been adopted for
investigating the FQHE at filling fractions with
odd-denominators at zero temperature in \cite{LF,SH,HSH},
and at finite temperature in \cite{LZ}.
For the $\nu =1/2$ state, in which the mean statistical
field and external magnetic field precisely cancel each other
such that the mean-field theory is just a filled Fermi sea,
this approach \cite{HLR} has successfully explained the
anomaly in the surface acoustic wave propagation
\cite{Willett}. Its qualitative correctness has also been
verified by a series of experiments \cite{CFexpt}.

Since the FQHE is observed in the presence of high
magnetic field $B$, it is conceivable that the FQHE may
be understood at the infinite magnetic field limit, where
the Hilbert space is composed solely by the states in
the lowest LL (LLL). This limit is important conceptually,
since in which the kinetic energy is frozen, showing clearly
the non-perturbative effect of electron-electron interactions.
Because the size of the Hilbert space is drastically reduced
by the restriction to the LLL, this limit is also of great
convenience for numerical studies which have been indispensable
to the development of our understanding of the FQHE.
The seminal work of Laughlin \cite{Laughlin} is formulated
entirely within the LLL. Noticeably that Jain's wavefunction,
whose construction is aided with states in higher LLs (to
utilize the mapping between FQH states and IQH states), also
needs to be ultimately mapped back to the LLL in order to give
the correct low energy physics \cite{Jain}.

The importance of the LLL is not explicit in the present
field theoretical approach, at least not in the first order
loop expansion commonly adopted in the literature. In fact
the electron-electron interaction, which is crucial for
the occurrence of the FQHE, plays only a nominal role in
the Chern-Simons field theory scheme mentioned above.
Formally, if one actually carried out the same calculation
for non-interacting electrons at these special fillings
(become clear later), one would obtain qualitatively the
same physics (e.g., incompressibility of the ground state)
as that with interactions. On the positive side, this fact
may be interpreted as the ``correctness" of the mean-field
theory \cite{RJ}; on the other hand, it is also quite
disturbing since we know that the ground state of a system
of non-interacting electrons with a partially filled LL is
compressible and is highly degenerate, and the FQHE occurs only
as a result of strong electron-electron correlations. Because
the interaction has not been treated properly, the energy
scale of the problem in this approach is incorrectly given
by an effective cyclotron energy which is on the same order
of the bare cyclotron energy, while the physical energy
scale is set by the electron-electron interaction.
This serious defect of the theory not only renders any
quantitative description of the low energy physics of the
FQHE impossible, but has also raised a fundamental question
concerning the validity of the theory itself.
In \cite{HLR,SH}, this problem
was dealt phenomenologically through a Fermi liquid theory.
In this approach, the magnetic field is treated semi-classically,
and the effective mass is introduced as a free parameter. This
Fermi liquid approach has been examined recently through a
comparison with exact numerical results on small systems \cite{HSH}.
While it provides a way of dealing the problem of mass
renormalization, its phenomenological nature is unsatisfactory.
Furthermore, for FQH states at filling fraction such as $\nu = 1/3$,
$1/5$, $2/5$ etc. in which the number of filled effective LLs
are small, the semi-classical approximation is expected to
break down due to the strong effective magnetic field.
What we would like to have is a microscopic formalism with a
mean-field theory for which minimal fluctuation corrections
are needed to reveal the qualitatively (and quantitatively)
correct physics.  With this philosophy in mind, we propose here
a fermion Chern-Simons field theory constrained to the LLL, in
which the essential role of the electron-electron interaction for
the FQHE is explicit. Our hope is that true physics may be
revealed in some low order perturbation expansion.

In the LLL, the kinetic energy is frozen and the effect
of interaction is non-perturbative. There are several works
in the literature dealing with the issue of formulating field
theories constrained in the LLL \cite{LLL1,LLL2}. However,
since the states in the LLL are incomplete, a field theory
constructed directly from the states in the LLL possesses
some peculiar properties which make evaluations of physical
properties extremely difficult. Here we shall follow an alternative
formalism. To illustrate our approach, it is instructive
to have a closer look at Jain's construction of the ground state
wavefunction for a FQH state \cite{Jain}. While the wavefunction
itself is built solely from the states in the LLL, it is
constructed in such a way that the role of higher LLs is
explicit. Mathematically, it provides a compact way of expressing
otherwise (i.e., writing directly within the Hilbert space of
the LLL) an extremely complicated object. Physically, it helps
tremendously in gaining insight (i.e., bridging the connection
between the FQHE and the IQHE) to the problem.
Our approach here is similar in spirit: instead of working
directly in the subspace spanned by the states in the LLL,
we shall start with the whole Hilbert space to utilize
the notion of mapping the FQHE onto the IQHE. The restriction
to the LLL is enforced by a $\delta$-functional described by
an auxiliary field $\lambda(r)$ (see below). Thus, just
like that in Jain's theory one employs states in the higher
LLs to construct the wavefunction but ultimately projects
it back to the LLL, here we use the
higher LLs in intermediate stages of calculations
and integrate out the $\lambda$-field at the end to realize
the constraint. As we shall see later, in this theory the
density and current fluctuations not only couples to themselves
and to each other, but also couples to a local kinetic energy
density such that the constraint to the LLL is satisfied.

A direct consequence of this approach is to provide a microscopic
calculation of the energy gap, which is experimentally measurable
\cite{book,Du,Sohn} and is an important quantity for estimating the effect
of thermal fluctuations on the accuracy of the FQHE. One may also
study the low energy collective excitation spectrum within
the present theory. The properties of low energy excitations may
be extrapolated from numerical results obtained through
exact diagonalization on small systems \cite{book} and from direct
construction of trial wavefunctions for low energy excited states
\cite{Jain,book,Laughlin,GMP}, or be studied through the
Chern-Simons-Landau-Ginzburg field theory \cite{CSLG}.
While these studies are very useful to our understanding
of the physics of the FQHE, for quantitative comparisons
with experiments we still need a {\sl microscopic} calculation
done directly in the {\sl thermodynamic limit}, especially in view
of the current discrepancies between experiments and theories
on the physical quantities such as the magnitude of the energy gap.
The previous works using the unconstrained fermion Chern-Simons
field theory \cite{LF,LZ} cannot serve this purpose
owing to its intrinsic difficulty with the energy scale
discussed above.

In the present work, we shall consider only those systems with
filling fraction $\nu=n/(1+2pn)$ ($p,n$ integer) whose
ground states are known to be incompressible and for whom there
are considerable amount of numerical results available which may
be helpful for testing the correctness of our theory. On the
other hand, computations, especially the numerical solution for
the dressed stationary point equation of the $\lambda$-field
in such a system are unfortunately much more complicated
than that for a $\nu =1/2p$ system in which the mean-field
ground state is just a filled fermi sea. However, since the
calculation of the $\nu=1/2p$ state is quite different from the
present one, it shall be studied in a separate work.

The rest of the paper is organized as following: in the next
section we present the formalism of our approach; in section III, we
study the electro-magnetic response; we discuss some consequences
of the theory and conclude in the section IV.

\section{Formalism}

The quantum many-body wave function describing a system of 2D
electrons of band mass $m_b$ in a magnetic field $\vec{B} =
{\vec \nabla} \times \vec{A}$ occupying only the LLL satisfies
the condition
\begin{equation}\label{con1}
{1\over{2m_b}}(-i\hbar{\vec \nabla}_{j}+\frac{e}{c}{\vec A})^{2}
\psi (\vec{r}_{1},...,\vec{r}_{j},...,\vec{r}_{N})
=\frac{1}{2}\hbar\omega_{c}
\psi (\vec{r}_{1},...,\vec{r}_{j},...,\vec{r}_{N})
\;\;,  \;\;\;\; \forall j \;,
\end{equation}
where $\omega_{c}=eB/m_{b}c$ is the cyclotron frequency.
In the second quantized language, one may express the
partition function for such a system by a coherent functional
integral subjected to a constraint:
\begin{equation}
{\cal Z} = \int {\cal D}\psi^{\dagger}{\cal D}\psi
{\cal D}\lambda e^{-S} \;\;,
\end{equation}
with the action (in the unit $\hbar=1$)
\begin{eqnarray}\label{action0}
S & = & \int_{0}^{\beta}d\tau\int d^{2}r \left \{
\psi^{\dagger}({\partial_\tau}-\mu)\psi(r)+ {1\over{2m_b}}
|(-i{\vec \nabla}+\frac{e}{c}{\vec A})\psi(r)|^{2}
\right \} + \nonumber \\
& & {1\over 2}\int_{0}^{\beta}d\tau \int d^{2}r\int d^{2}r^{\prime}
\psi^{\dagger}\psi(r) v(\vec{r}-\vec{r}\;^{\prime})
\psi^{\dagger}\psi(r^{\prime}) +
\nonumber \\
& & i\int_{0}^{\beta}d\tau\int d^{2}r \lambda(r) \left \{ {1\over{4m_b}}
\left [ \psi^{\dagger}(-i{\vec \nabla}+\frac{e}{c}{\vec A})^{2}\psi(r)
+((i{\vec \nabla}+\frac{e}{c}{\vec A})^{2}\psi^{\dagger})
\psi(r) \right ] -\frac{1}{2}\omega_{c}\psi^{\dagger}\psi(r)
\right \} \;.
\end{eqnarray}
In the above expression,
$\mu$ is the chemical potential which fixes the LL
filling fraction at $\nu$. $v(\vec{r})$ is a two-body interaction
potential. In this paper we
shall consider the Coulomb interaction explicitly in our
calculation, where $v(\vec{r})=e^{2}/\epsilon r$, and $\epsilon$
is the dielectric constant. Other forms of interaction may be
substituted straightforwardly into our calculations.
The last term is a $\delta$-functional enforcing the constraint
(\ref{con1}), where the kinetic energy term is symmetrized with
respect to $\psi^{\dagger}$ and $\psi$.
Notice that since the kinetic and potential energy term do not
commute with each other, the constraint is intrinsically dynamical.
Thus $\lambda(r) \equiv \lambda(\vec{r},\tau)$ depends explicitly
on the (imaginary) time $\tau$. This is different from that in the
situation in, e.g., the slave boson approach for lattice fermions
with infinite on-site repulsion in which the no-double-occupancy
constraint may also be enforced by a $\delta$-functional \cite{SB}.
The constraint equation there is a constant
of motion hence $\lambda(r) = \lambda(\vec{r})$ is time independent.
Because of the local nature of the auxiliary field $\lambda$, the
kinetic energy in the last term may not be rearranged into the usual
form ${1\over{2m_b}}|(-i{\vec \nabla}+\frac{e}{c}{\vec A})\psi|^{2}$
through partial integration. In the spirit of composite fermion
approach, we rewrite the action (\ref{action0}) in terms of the
``transformed fermion fields", still denoted by $\psi^\dagger$ and
$\psi$, such that
\begin{equation}
{\cal Z} = \int {\cal D}\psi^{\dagger}{\cal D}\psi
{\cal D} a_{\mu} {\cal D}\lambda {e}^{-S} \;\;,
\end{equation}
with
\begin{eqnarray}\label{action1}
\lefteqn{
S=\int_{0}^{\beta}d\tau\int d^{2}r \left \{
\psi^{\dagger}({\partial_\tau}-ia_{0}-\mu)\psi+
{1\over{2m_b}}|(-i{\vec \nabla}+\frac{e}{c}{\vec A}-{\vec a})\psi|^{2}
+{i\over {4\pi\tilde{\phi}}}\epsilon_{\mu\nu\lambda}
a_{\mu}{\partial}_{\nu}a_{\lambda} \right \} +
} \nonumber \\
& & {1\over 2}\int_{0}^{\beta}d\tau \int d^{2}r\int d^{2}r^{\prime}
\psi^{\dagger}(r)\psi(r) v(\vec{r}-\vec{r}\;^{\prime})
\psi^{\dagger}(r^{\prime})\psi(r^{\prime}) + \nonumber \\
& & i\int_{0}^{\beta}d\tau\int d^{2}r \lambda(r) \left \{ {1\over{4m_b}}
\left [ \psi^{\dagger}(-i{\vec \nabla}+
\frac{e}{c}{\vec A}-{\vec a})^{2}\psi(r) + ((i{\vec \nabla}+
\frac{e}{c}{\vec A}-{\vec a})^{2}\psi^{\dagger}) \psi(r) \right ]
-\frac{1}{2}\omega_{c}\psi^{\dagger}(r)\psi(r)  \right \} \;\;.
\end{eqnarray}
Here ${\tilde \phi} = 2p$, and $p$ is an integer. In the integral
of $a_{\mu}$, a gauge fixing procedure is implicitly assumed.
We attached even ($2p$) flux quanta per fermion through the
well-known ``Chern-Simons term" with statistical gauge field
$a_\mu$: integrating out the $0$-component of $a_\mu$, we obtain
\begin{equation}\label{constraint}
|\vec{\nabla} \times \vec{a}| = 2\pi\tilde{\phi}\rho(r) \;\;,
\end{equation}
where $\rho(r)=\psi^{\dagger}\psi(r)$ is a local density.
While in the rest of the paper we will mainly focused on
the ground state properties, we have formulated our theory
within a finite temperature field theory formalism.  Similar
actions but without the constraint have been investigated
in \cite{RSS} in the context of anyon superconductivity
and in \cite{LZ} in a study of the FQHE at finite temperature.

Since the ground state of free fermions in a partially
filled LL is highly degenerate, it is clearly not suitable
for perturbative expansions.
The above gauge transformation is motivated by the desire of
finding a suitable mean-field theory which may provide a
convenient starting point for approximations. Without the
constraint and at zero $T$,
this has been investigated by Lopez and Fradkin \cite{LF}:
consider the homogeneous liquid saddle point solution for
the Chern-Simons field $a_{\mu}=\bar{a}_{\mu}$, such that
\begin{equation}\label{mfcs}
|\vec{\nabla} \times \vec{\bar{a}}| = 2\pi\tilde{\phi}\bar{\rho}
\;\;, \;\;\;\;\;\; \bar{a}_{0} = 0 \;\;,
\end{equation}
where $\bar{\rho}$ is the average particle density. In this
mean-field theory, the single-particle orbitals are described
by the LLs determined by the effective magnetic field
\begin{equation}\label{Beff}
B_{eff} \equiv |\vec{\nabla} \times \vec{A}_{eff}| \equiv
|\vec{\nabla} \times (\vec{A} - \frac{c}{e}\vec{\bar{a}})|
\;\;\;.
\end{equation}
For the filling fraction $\nu$ such that $1/(1/\nu-2p) = n$,
an integer, $B_{eff}= \nu B/n$, the mean-field theory possesses
an incompressible ground state described by $n$ filled (effective)
LLs. The filling fraction $\nu$ which satisfies this condition is
precisely the value where the FQHE arises. However, since this
mean-field approximation for the Chern-Simons statistical gauge
field $a_\mu$ does not depend on the nature of the underlying
electronic state (other than the assumption that it is a
homogeneous liquid), in the {\em unconstrained} theory this
mapping of such a fractionally filled state onto an
incompressible state also works (by which we mean that the
mean-field solution exists and is stable against Gaussian
fluctuations) for free electrons!
This contradicts with the well known fact that the
FQHE is a result of strong electron-electron correlations hence
the mapping between the FQH states and the IQH states is sensible
only when interactions are present. This serious deficiency of
the unconstrained theory is reflected by the fact that the energy
scale describing the FQH states in this theory is given the
effective cyclotron energy $\hbar eB_{eff}/m_{b}c$, which is
proportional to the bare cyclotron energy $\hbar\omega_{c}$ \cite{LF}.

Now we consider the homogeneous mean-field solution {\em with} the
constraint. The mean-field theory is obtained by substitutions
$a_{\mu} \rightarrow \bar{a}_{\mu}$, $i\lambda(r) \rightarrow
\bar{\lambda}$, where $\bar{a}_{\mu}$ is given by Eqn.(\ref{mfcs})
and $\bar{\lambda}$ is a constant, independent of $r$, such that
$\partial {\cal F}/\partial \bar{\lambda} =0$ enforces the LLL
constraint. In the free electron case, the ground state
of the mean-field theory for the statistical field (\ref{mfcs})
is given by the configurations of electrons occupying the
effective LLs subjected to the constraint
\begin{equation}\label{meanE}
\sum_{l}\hbar\bar{\omega}_{c}(l+\frac{1}{2})n_{l}
=\frac{1}{2}\hbar\omega_{c} N_{e} \;\;,
\end{equation}
where $\bar{\omega}_{c}=eB_{eff}/m_{b}c$ is the mean-field
cyclotron frequency and $n_l$ is the occupancy of the l-th LL.
Consider the example of $\nu=1/3$. In this case $\bar{\omega}_{c}
=\frac{1}{3}\omega_{c}$, and the mean-field eigenspectrum is given
by $\frac{\hbar\omega_c}{6}(l+\frac{1}{2})$. As illustrated in
Fig. 1, this ground state is highly degenerate as expected.
Thus we see that there is indeed no resemblance of a fractionally
filled state to the IQH state for free systems.
Notice that in this (non-interacting) case, the mean-field
value for the $\lambda$-field $i\lambda(r) = \bar{\lambda}$
is not a stationary point since the free energy is a linear
function of $\bar{\lambda}$.

Consider now the interacting case. As shown in the Ref. \cite{LF,HLR},
the treatment of interaction is equivalent to a treatment of the
fluctuations in the gauge field $a_{\mu}$ because of the condition
(\ref{constraint}). Thus one may write the interaction term in the
action (\ref{action1}) in terms of the statistical field $a_{\mu}$.
Shift
\begin{equation}
a_{\mu} \rightarrow \bar{a}_{\mu} +a_{\mu}\;\;, \;\;\;\;
i\lambda(r) \rightarrow \bar{\lambda} +i\lambda(r) \;\;,
\end{equation}
so that $a_\mu$ and $\lambda$ represent fluctuations with
respect to the saddle point solution, we write the action
$S$ in the form $S=S_{0}+S_{1}$, with
\begin{equation}\label{MFaction}
S_{0}= \int_{0}^{\beta}d\tau\int d^{2}r \left \{ \psi^{\dagger}
({\partial_\tau}-\tilde{\mu})\psi(r)+ {1\over{2m^{*}}}
|(-i{\vec \nabla}+\frac{e}{c}{\vec A_{eff}})\psi(r)|^{2}
\right \} \;\;,
\end{equation}
and
\begin{eqnarray}
S_{1} & = & \int_{0}^{\beta}d\tau\int d^{2}r \left \{ (-ia_{0}
-i\lambda(r)\frac{\omega_c}{2})\rho(r)
-\vec{j}\cdot \vec{a}(r)+\frac{1}{2m^{*}}\vec{a}\cdot \vec{a}\rho(r)
+{i\over {4\pi\tilde{\phi}}}\epsilon_{\mu\nu\lambda}
a_{\mu}{\partial}_{\nu}a_{\lambda} \right \} \nonumber \\
& & +i\lambda(r)\frac{m^{*}}{m_{b}}h(r)
-i\lambda(r)\frac{m^{*}}{m_{b}}\vec{j}\cdot \vec{a}(r)
+i\lambda(r)\frac{1}{2m_b}\vec{a}\cdot \vec{a}\rho(r)
\nonumber \\
& & +{1\over 2}\int_{0}^{\beta}d\tau \int d^{2}r\int d^{2}r^{\prime}
{1\over (2\pi\tilde{\phi})^2} [{\vec \nabla} \times \vec{a}(r)]
v(\vec{r}-\vec{r}) [{\vec \nabla} \times \vec{a}(r^{\prime})] \;\;,
\end{eqnarray}
where
\begin{equation}
m^{*}=m/(1+\bar{\lambda}) \;\;,
\end{equation}
\begin{equation}\label{renormu}
\tilde{\mu}=\mu + \frac{\bar{\lambda}}{2}\omega_{c} \;\;,
\end{equation}
are renormalized mass and renormalized chemical potential respectively.
\begin{equation}\label{mfcd}
\vec{j}(r)=-\frac{i}{2m^{*}}(\psi^{\dagger}{\vec \nabla}\psi(r)-
({\vec \nabla}\psi^{\dagger})\psi(r)) +
\frac{1}{m^{*}}\frac{e}{c}{\vec A_{eff}}\psi^{\dagger}\psi(r)
\end{equation}
is the mean-field current density and
\begin{equation}
h(r)=\frac{1}{4m^{*}} \left [ \psi^{\dagger}
(-i{\vec \nabla}+\frac{e}{c}{\vec A}_{eff})^{2}\psi(r) +
((i{\vec \nabla}+\frac{e}{c}{\vec A}_{eff})^{2}\psi^{\dagger})
\psi(r)  \right ]
\end{equation}
is the mean-field kinetic energy density.
$\bar{\lambda}$ is determined by the extremal condition
\begin{equation}\label{dfdlambda}
\frac{\partial \cal{F}}{\partial \overline{\lambda}} = 0 \;\;.
\end{equation}
Without $S_{1}$, the mean-field ground state is highly degenerate
as discussed above. Following the ``composite fermion" picture
\cite{Jain}, it is nature to expect that a new
stationary point solution, which describes $n$ filled (effective)
LLs for the filling fraction $\nu= n/(1+2pn)$, will emerge when the
interaction is properly treated.  However, it is a non-trivial task to
determine which level of approximations would qualify to be the ``proper
treatment" of the interaction. In the present work, we will only examine
the single loop quantum corrections described by a Gaussian theory.

As the very first step for constructing a loop expansion,
one needs to decide which mean-field state should be taken.
Based on the above discussion, we shall assume that fluctuations
will shift the non-interacting mean-field ground state
to a new one which is non-degenerate and is described by $n$
filled effective LLs for the filling fractions $\nu=n/(1+2pn)$.
With this new mean-field state, we calculate the (one) loop
corrections and obtain the physical description of the
corresponding FQH state. The correctness of our assumption will be
determined later by a self-consistency requirement. As will be
discussed below, the self-consistency equation may be seen as
the equation for the mass renormalization of composite fermions.

The fact that fluctuation corrections may result in a new
stationary point state is well documented. For a charged
superconductor, it is known that in certain situations \cite{HLM}
gauge field fluctuations may shift the order parameter
and render the superconducting transition to first order,
in contrast to the usual second order phase transition predicted
by the Ginzburg-Landau theory without considering the fluctuations
in the electric-magnetic field \cite{SCbook}.
In quantum field theory, a similar case is also found
in the scalar electrodynamics in four dimension \cite{CW}.
In our case, the situation is slightly different since here
we postulate the new mean-field solution first, compute the
loop corrections with respect to this assumed mean-field
solution, and verify the correctness (i.e., the existence
and the stability of this new solution) self-consistently
by solving the saddle point equation, which is now dressed by
loop corrections. Our hope is that the one-loop theory will
be sufficient to give the correct physics.
An idea of similar spirit has been proposed previously by Weng
in his treatment of doped quantum antiferromagnet \cite{Weng}.

If one ignores the fluctuations in $\lambda$, and considers
only the Gaussian fluctuations in the statistical field
$a_\mu$, one would obtain similar results as those of the
unconstrained theory \cite{LF,LZ}, but with a renormalized
mass $m^*$. However, this simple minded approach violates
Kohn's theorem \cite{Kohn}. Thus it is essential to incorporate
the fluctuations in $\lambda$, treat them at the same footing
as those in the statistical field $a_\mu$.

As noted in \cite{HLR}, in a theory without constraint,
one may utilize the continuity equation to simplify the
calculation of the response kernel through some special
choice of gauge. For a system of electrons constrained to
the LLL, however, it is in general not possible to apply
such a procedure since in this case the current defined
in the usual manner does not satisfy the continuity equation
\cite{LLL1,LLL2}. Thus we shall use the gauge invariant
formalism of \cite{LF}. Denote $(a_{\mu} \; i\lambda) \equiv
(ia_{0} \; a_{x} \; a_{y} \; i\lambda)$, the Gaussian action
may be expressed as
\begin{eqnarray}
S_{Gaussian}(a_{\mu}; i\lambda)= \frac{1}{2}\sum_{\vec{q},i\omega_{n}}
\left ( a_{\mu}(-\vec{q},-i\omega_{n})\;\;\;
i\lambda(-\vec{q},-i\omega_{n})\right)
D^{-1}(\vec{q},i\omega_{n})
\left ( \begin{array}{c} a_{\mu}(\vec{q},i\omega_{n}) \\
i\lambda(\vec{q},i\omega_{n})\end{array} \right ) \;\;,
\end{eqnarray}
with the kernel
\begin{equation}
D^{-1}(\vec{q},i\omega_{n})=U(\vec{q},i\omega_{n}) +
\Pi^{0}(\vec{q},i\omega_{n}) \;\;.
\end{equation}
Here $U$ is the tree diagram contribution
\begin{equation}
U(\vec{q},i\omega_{n})=\frac{1}{2\pi\tilde{\phi}}\left(
\begin{array}{cccc}
0&-iq_{y}&iq_{x} &0\\
iq_{y}&\tilde{\omega}_{c}\frac{\mu_s}{2p}\frac{q_{y}^{2}}{q^2}
\tilde{l}_{0}q&i(i\omega_{n})-
\tilde{\omega}_{c}\frac{\mu_s}{2p}\frac{q_{x}q_{y}}{q^2}
\tilde{l}_{0}q &0\\
-iq_{x}&-i(i\omega_{n})- \tilde{\omega}_{c}\frac{\mu_s}{2p}
\frac{q_{x}q_{y}}{q^2}\tilde{l}_{0}q& \tilde{\omega}_{c}
\frac{\mu_s}{2p}\frac{q_{x}^{2}}{q^2} \tilde{l}_{0}q &0 \\
0&0&0&0
\end{array}\right)
\end{equation}
and $\Pi^{0}(\vec{q}, i\omega_{n})$ gives the one-loop contribution
(for $i\omega_{n} \neq 0$),
\begin{eqnarray}\label{Pi0}
\lefteqn {\Pi^{0}(\vec{q},i\omega_{n})  = } \nonumber \\
& \frac{\tilde{n}}{2\pi\tilde{\phi}} \left(
\begin{array}{cccc}
\frac{q^{2}}{\tilde{\omega}_{c}}\Sigma_{0}&
\frac{i\omega_n}{\tilde{\omega}_c}q_{x}\Sigma_{0}+iq_{y}\Sigma_{1}&
-iq_{x}\Sigma_{1}+\frac{i\omega_n}{\tilde{\omega}_c}q_{y}\Sigma_{0}&
\frac{\omega_{c}}{2}\frac{q^2}{\tilde{\omega}_{c}}\Sigma_{4} \\
\frac{i\omega_n}{\tilde{\omega}_c}q_{x}\Sigma_{0}-iq_{y}\Sigma_{1}&
\tilde{\omega}_{c}((\frac{i\omega_n}{\tilde{\omega}_c})^{2}\Sigma_{0}
+\frac{q_{y}^{2}}{q^2}\Sigma_{3})& -i(i\omega_{n})\Sigma_{1}
-\tilde{\omega}_{c}\frac{q_{x}q_{y}}{q^2}\Sigma_{3}&
\frac{\omega_{c}}{2}(\frac{i\omega_n}{\tilde{\omega}_c}q_{x}\Sigma_{4}
-iq_{y}\Sigma_{5}) \\
iq_{x}\Sigma_{1}+\frac{i\omega_n}{\tilde{\omega}_c}q_{y}\Sigma_{0}&
i(i\omega_{n})\Sigma_{1}-
\tilde{\omega}_{c}\frac{q_{x}q_{y}}{q^2}\Sigma_{3}&
\tilde{\omega}_{c}((\frac{i\omega_n}{\tilde{\omega}_c})^{2}\Sigma_{0}
+\frac{q_{x}^{2}}{q^2}\Sigma_{3}) & \frac{\omega_{c}}{2}(
iq_{x}\Sigma_{5}+\frac{i\omega_n}{\tilde{\omega}_c}q_{y}\Sigma_{4}) \\
\frac{\omega_{c}}{2}\frac{q^2}{\tilde{\omega}_{c}}\Sigma_{4} &
\frac{\omega_{c}}{2}(\frac{i\omega_n}{\tilde{\omega}_c}q_{x}\Sigma_{4}
+iq_{y}\Sigma_{5}) & \frac{\omega_{c}}{2}(-iq_{x}\Sigma_{5}+
\frac{i\omega_n}{\tilde{\omega}_c}q_{y}\Sigma_{4}) &
(\frac{\omega_{c}}{2})^{2}\frac{q^2}{\tilde{\omega}_c}\Sigma_{6}
\end{array}\right) \;\;,
\end{eqnarray}
with $\tilde{n}=2pn$ and
\begin{eqnarray}
\Sigma_{j} & = & \frac{e^{-x}}{n}\sum_{m<l}
\frac{l-m}{(\frac{i\omega_n}{\tilde{\omega}_{c}})^{2}-(l-m)^{2}}
\frac{m!}{l!}x^{l-m-1}
\{ f(\epsilon_{m}-\tilde\mu)-f(\epsilon_{l}-\tilde\mu) \} \nonumber \\
& & \left [ L_{m}^{l-m}(x) \right ]^{2-j} \left [
(l-m-x)L_{m}^{l-m}(x)+2x\frac{dL_{m}^{l-m}(x)}{dx} \right ] ^{j}
\end{eqnarray}
for j=0,1,2 \cite{fnote2}; and
\begin{eqnarray}
\Sigma_{3} & = & \frac{e^{-x}}{n}\sum_{m<l}
\frac{l-m}{(\frac{i\omega_n}{\tilde{\omega}_{c}})^{2}-(l-m)^{2}}
\frac{m!}{l!}x^{l-m}
\{ f(\epsilon_{m}-\tilde\mu)-f(\epsilon_{l}-\tilde\mu) \} \nonumber \\
& & \left [ 2\frac{dL_{m}^{l-m}(x)}{dx} - L_{m}^{l-m}(x) \right ]
\left [ (2(l-m)-x)L_{m}^{l-m}(x)+2x\frac{dL_{m}^{l-m}(x)}{dx} \right ]
\;\;,
\end{eqnarray}
\begin{eqnarray}
\Sigma_{4}=\frac{e^{-x}}{n}\sum_{m<l} \frac{l-m}
{(\frac{i\omega_n}{\tilde{\omega}_{c}})^{2}-(l-m)^{2}} \frac{m!}{l!}
x^{l-m-1} \{ f(\epsilon_{m}-\tilde\mu)-f(\epsilon_{l}-\tilde\mu) \}
\left (1-\frac{l+m+1}{1+\tilde{n}} \right )
\left [ L_{m}^{l-m}(x) \right ]^{2} \;,
\end{eqnarray}
\begin{eqnarray}
\Sigma_{5} & = & \frac{e^{-x}}{n}\sum_{m<l}
\frac{l-m}{(\frac{i\omega_n}{\tilde{\omega}_{c}})^{2}-(l-m)^{2}}
\frac{m!}{l!}x^{l-m-1}
\{ f(\epsilon_{m}-\tilde\mu)-f(\epsilon_{l}-\tilde\mu) \} \nonumber \\
& & \left ( 1-\frac{l+m+1}{1+\tilde{n}} \right ) L_{m}^{l-m}(x)
\left [ (l-m-x)L_{m}^{l-m}(x)+2x\frac{dL_{m}^{l-m}(x)}{dx} \right ]
\;\;,
\end{eqnarray}
\begin{eqnarray}
\Sigma_{6}=\frac{e^{-x}}{n}\sum_{m<l}
\frac{l-m}{(\frac{i\omega_n}{\tilde{\omega}_{c}})^{2}-(l-m)^{2}}
\frac{m!}{l!}x^{l-m-1}
\{ f(\epsilon_{m}-\tilde\mu)-f(\epsilon_{l}-\tilde\mu) \}
\left ( 1-\frac{l+m+1}{1+\tilde{n}} \right )^2
\left [ L_{m}^{l-m}(x) \right ]^{2} \;.
\end{eqnarray}
In the above equations, $i\omega_{n}$ is the Matsubara frequency,
$L_{m}^{l}$ is the Laguerre polynomial, $x=(\tilde{l}_{0}q)^{2}/2$,
and $\tilde{l}_{0}=\sqrt{ c/eB_{eff}}$,
$\tilde{\omega}_{c}=eB_{eff}/m^{*}c$, are the effective magnetic
length and the effective cyclotron frequency respectively.
$\mu_{s}$ is the ratio of $\tilde{l}_{0}$ to the effective Bohr
radius $a_{0}=\epsilon /m^{*}e^{2}$. The chemical potential $\mu$
is renormalized by $\bar{\lambda}$ according to (\ref{renormu})
such that there are precisely $n$ LL are filled \cite{RSS}:
\begin{equation}\label{chemp}
n=\sum_{l}f(\epsilon_{l}-\tilde{\mu})\;\;, \;\;\;\;\;
f(\epsilon_{l}-\tilde{\mu}) =
\frac{1}{e^{\beta(\epsilon_{l}-\tilde{\mu})} + 1} \;\;\;,
\end{equation}
where $\epsilon_{l}=\tilde{\omega}_{c}(l+\frac{1}{2})$.
For the case $i\omega_{n} = 0$, the one-loop contribution is given by
\begin{equation}\label{Pi00}
\Pi^{0}(\vec{q},i\omega_{n}=0)=
\frac{\tilde{n}}{2\pi\tilde{\phi}}\left(
\begin{array}{cccc}
\frac{q^{2}}{\tilde{\omega}_{c}}\Sigma_{0}^{0}&
iq_{y}\Sigma_{1}^{0}& -iq_{x}\Sigma_{1}^{0}&
\frac{\omega_{c}}{2}\frac{q^2}{\tilde{\omega}_{c}}\Sigma_{4}^{0} \\
-iq_{y}\Sigma_{1}^{0}&
\tilde{\omega}_{c}\frac{q_{y}^{2}}{q^2}\Sigma_{3}^{0}&
-\tilde{\omega}_{c}\frac{q_{x}q_{y}}{q^2}\Sigma_{3}^{0}&
-i\frac{\omega_{c}}{2} q_{y}\Sigma_{5}^{0} \\
iq_{x}\Sigma_{1}^{0}&
-\tilde{\omega}_{c}\frac{q_{x}q_{y}}{q^2}\Sigma_{3}^{0}&
\tilde{\omega}_{c}\frac{q_{x}^{2}}{q^2}\Sigma_{3}^{0} &
i\frac{\omega_{c}}{2} q_{x}\Sigma_{5}^{0} \\
\frac{\omega_{c}}{2}\frac{q^2}{\tilde{\omega}_{c}}\Sigma_{4}^{0} &
i\frac{\omega_{c}}{2} q_{y}\Sigma_{5}^{0} &
-i\frac{\omega_{c}}{2} q_{x}\Sigma_{5}^{0} &
(\frac{\omega_{c}}{2})^{2}\frac{q^2}{\tilde{\omega}_c}\Sigma_{6}^{0}
\end{array}\right) \;\;\;,
\end{equation}
where
\begin{equation}
\Sigma_{j}^{0}(\vec{q})=\Sigma_{j}(\vec{q},0)-
\frac{\beta\tilde{\omega}_c}{2} \frac{e^{-x}}{n}x^{j-1}\sum_{m}
e^{(\epsilon_{m}-\tilde\mu)}f^{2}(\epsilon_{m}-\tilde\mu)
\left [ L_{m}^{0}(x) \right ]^{2-j}
\left [ (L_{m}^{0}(x)-2L_{m}^{1}(x) \right ] ^{j}
\end{equation}
for j=0,1,2;
\begin{equation}
\Sigma_{3}^{0}(\vec{q})=\Sigma_{2}^{0}(\vec{q}) \;\;,
\end{equation}
\begin{equation}
\Sigma_{4}^{0}(\vec{q})=\Sigma_{4}(\vec{q},0)-
\frac{\beta\tilde{\omega}_c}{2}
\frac{e^{-x}}{nx}\sum_{m}\left (1-\frac{2m+1}{1+\tilde{n}} \right )
e^{(\epsilon_{m}-\tilde\mu)}f^{2}(\epsilon_{m}-\tilde\mu)
\left [ L_{m}^{0}(x) \right ]^{2} \;\;,
\end{equation}
\begin{equation}
\Sigma_{5}^{0}(\vec{q})=\Sigma_{5}(\vec{q},0)-
\frac{\beta\tilde{\omega}_c}{2}
\frac{e^{-x}}{n}\sum_{m}\left (1-\frac{2m+1}{1+\tilde{n}} \right )
e^{(\epsilon_{m}-\tilde\mu)}f^{2}(\epsilon_{m}-\tilde\mu)
L_{m}^{0}(x) \left [ L_{m}^{0}(x)-2L_{m}^{1}(x) \right ]  \;,
\end{equation}
\begin{equation}
\Sigma_{6}^{0}(\vec{q})=\Sigma_{6}(\vec{q},0)-
\frac{\beta\tilde{\omega}_c}{2} \frac{e^{-x}}{nx}\sum_{m}
\left [1-\frac{2m+1}{1+\tilde{n}}\right ]^{2}
e^{(\epsilon_{m}-\tilde\mu)}f^{2}(\epsilon_{m}-\tilde\mu)
\left [ L_{m}^{0}(x) \right ] ^{2} \;\;.
\end{equation}
The computation for $\Pi^0$ is detailed in the appendix.
Notice that because there are no tree diagrams involving
the auxiliary field $i\lambda(r)$, the matrix $U$ possesses
no inverse.  The free energy is obtained by completing the
functional integral (with the standard gauge fixing procedure)
of $a_{\mu}$ and $\lambda$ in the partition function:
\begin{equation}
{\cal F} = {\cal F}_{0}+\frac{1}{2\beta}
\sum_{\vec{q},i\omega_n} \ln det D^{-1}(\vec{q},i\omega_{n}) \;\;,
\end{equation}
where ${\cal F}_0$ is the mean-field contribution to the free
energy and the determinant of $D$, $det D$, is given by, apart
from a constant (in $\bar\lambda$) factor (see \cite{fnote2}),
\begin{eqnarray}\label{detD}
det D  & = &
\tilde{n}(\Sigma_{0}\Sigma_{6}-\Sigma_{4}^{2})
\frac{\mu_s}{2p}\tilde{l}_{0}q
+\tilde{n}^{2}(\Sigma_{0}\Sigma_{6}-\Sigma_{4}^{2})
[(\frac{i\omega_n}{\tilde{\omega}_c})^2\Sigma_{0}+\Sigma_{3}]
\nonumber \\
& & -\Sigma_{6}(1-\tilde{n}\Sigma_{1})^{2}-
\tilde{n}^{2}\Sigma_{0}\Sigma_{5}^{2}-
2\tilde{n}\Sigma_{4}\Sigma_{5}(1-\tilde{n}\Sigma_{1}) \;\;.
\end{eqnarray}
The saddle point value $\bar{\lambda}$ is given by the extremal
condition (\ref{dfdlambda}) which, in the present case of
one-loop expansion, can be written as
\begin{equation}\label{massrn}
\frac{\partial {\cal F}_0}{\partial \bar\lambda} +
\frac{1}{2\beta} \sum_{\vec{q},i\omega_n} \frac{1}
{det D(\vec{q},i\omega_{n})} \frac{\partial det
D(\vec{q},i\omega_{n})}{\partial {\bar \lambda}} =0 \;\;.
\end{equation}
If one lets $\mu_{s}=0$ in (\ref{detD}), which corresponds
to that for a non-interacting system with Gaussian
fluctuation corrections (in both $a_\mu$ and $\lambda$),
one would easily see that $\cal F$ is linear in $\bar\lambda$
and therefore there is no saddle point.
Because of the interaction, the free energy is no more a
linear function of $\bar{\lambda}$, hence one may expect a
non-trivial solution. Thus, after taking a mean-field state of
integer-filled pseudo LLs (which is not a legitimate state
for non-interacting electrons but is assumed to emerge as a
valid one when the interaction is switched on) as our starting
point for the loop expansion, our assumption is checked
self-consistently by finding solutions for (\ref{massrn}),
in which the free energy is calculated with the one-loop
fluctuation corrections.

{}From the expression of $S_{0}$, Eqn.(\ref{MFaction}), it is
tempting to interpret $m^{*} \equiv m/(1+\bar{\lambda})$
as an ``effective mass" of the composite fermions. (Notice
that the stability of the mean field solution requires that
$1+\bar{\lambda} >0$.)  Thus the above equation (\ref{massrn})
may be seen as the mass renormalization equation for composite
fermions. While nominally this definition of the effective mass
differs from the usual one \cite{HLR,KLWS}, it too is a result
of electron-electron interactions and gauge field fluctuations.
And we shall see later, $m^*$ also measures the energy gap in the
excitation spectrum.  Incorporating higher order loop expansions
will in general further shift $\bar{\lambda}$, giving additional
corrections to the effective mass.

\section{Electro-Magnetic Response}

The electro-magnetic response for a system constrained to the LLL
is a quite subtle problem. Consider fluctuations in $A_{\mu}$
such that $A_{\mu} \rightarrow A_{\mu} + \delta A_{\mu}$.
It is tempting to write the action describing the system with this
additional fluctuating electric-magnetic field as
\begin{eqnarray}\label{action2}
S & = & \int_{0}^{\beta}d\tau\int d^{2}r \left \{ \psi^{\dagger}
({\partial_\tau}+i\delta A_{0} -\mu)\psi(r) + {1\over{2m_b}}
|(-i{\vec \nabla}+\frac{e}{c}({\vec A}+\delta {\vec A}))\psi(r)|^{2}
\right \} + \nonumber \\
& & {1\over 2}\int_{0}^{\beta}d\tau \int d^{2}r\int d^{2}r^{\prime}
\psi^{\dagger}\psi(r) v(\vec{r}-\vec{r}\;^{\prime})
\psi^{\dagger}\psi(r^{\prime}) +
\nonumber \\
& & i\int_{0}^{\beta}d\tau\int d^{2}r \lambda(r) \left \{ {1\over{4m_b}}
\left [ \psi^{\dagger}(-i{\vec \nabla}+\frac{e}{c}{\vec A})^{2}\psi(r)
+((i{\vec \nabla}+\frac{e}{c}{\vec A})^{2}\psi^{\dagger})
\psi(r) \right ] -\frac{1}{2}\omega_{c}\psi^{\dagger}\psi(r)
\right \} \;.
\end{eqnarray}
Notice that the term involving the $\lambda$-field is independent
of $\delta A_{\mu}$. One would expect that such an action describes
the electro-magnetic response of the system to the perturbation
$\delta A_{\mu}$ while it is confined to the Hilbert space of the
LLL of the unperturbed system, consistent with the ``standard
approach" using trial wavefunctions \cite{book}. However, as it
stands this action is not invariant under the gauge transformation
\begin{equation}
\delta A_{\mu} \rightarrow \delta A_{\mu} + \partial_{\mu} f \;\;.
\end{equation}
Because one does not have the usual ``minimal coupling" for the
external perturbation $\delta A_\mu$, the current density
cannot be simply obtained from the derivative
\begin{equation}
j_{\mu} = \frac{\partial S}{\partial  A_{\mu}} \;\;.
\end{equation}
On the other hand, the
action (\ref{action0}) (or equivalently (\ref{action1})) clearly
preserves the global $U(1)$ symmetry of the unconstrained theory,
one may thus still define a Noether current density accordingly.
As usual, consider a phase change in the fermion field such that
\begin{equation}
\psi \rightarrow e^{i\epsilon}\psi \;,\;\;\;\;\;
\psi^{\dagger} \rightarrow \psi^{\dagger}e^{-i\epsilon} \;\;.
\end{equation}
Invariance of the action with respect to an arbitrary infinitesimal
change of $\epsilon$ lead to a conserved current density
$j^{\lambda}_{\mu}$, which in terms of the Chern-Simons
transformed fields is given by
\begin{eqnarray}
j^{\lambda}_{0} & =& \psi^{\dagger}\psi(r) \;\;,
 \nonumber \\
{\vec j}^{\lambda} & = & {\vec j}(r) - \frac{\vec a}{m^*}
\psi^{\dagger}\psi(r) + i\lambda(r) \left \{
-\frac{i}{2m_b} [\psi^{\dagger}{\vec \nabla}\psi(r)-
({\vec \nabla}\psi^{\dagger})\psi(r)] + \frac{1}{m_b}
(\frac{e}{c}{\vec A_{eff}}-{\vec a})\psi^{\dagger}\psi(r)
\right \}  \;,
\end{eqnarray}
where ${\vec j}(r)$ is the mean-field current density defined
in (\ref{mfcd}). $j^{\lambda}_{\mu}$ is, of course, not the
physical current density measured in experiments, which must
be obtained by integrating out the auxiliary field $\lambda$
\cite{LLL2}.

To study the electro-magnetic response, we shall consider a
simpler case where the external fluctuating field is in the
time component only, i.e., $\delta A_{\mu} = ( i\delta A_{0},
0,0)$, so that it couples to the system according to
(\ref{action0}) with the substitution
$A_{\mu} \rightarrow A_{\mu} + \delta A_{\mu}$.
The action is then invariant under the gauge transformation
$A_{\mu} +\delta A_{\mu} \rightarrow A_{\mu} + \delta A_{\mu} +
\partial_{\mu} f$, and the system under such a perturbation
$\delta A_{\mu}$ is still constrained to the (original) LLL.
Using the standard Faddeev-Popov procedure (where the gauge
fixing term can be chosen as, e.g.,
$\frac{1}{2\alpha}(\partial_{\mu} a_{\mu})^{2})$, one may
integrate out the statistical field $a_\mu$ and arrives an
effective action for such a $\delta A_{\mu}$. After some
algebra, one obtains
\begin{equation}
{\tilde S}_{Gaussian}(\delta A_{0}) =
\frac{1}{2}\sum_{\vec{q},i\omega_{n}}
i\delta A_{0}(-\vec{q},-i\omega_{n})\;
K_{00}(\vec{q},i\omega_{n}) \; i\delta A_{0}(\vec{q},i\omega_{n}) \;\;.
\end{equation}
One may interpret $K_{00}$ as the density-density
correlation function \cite{LLL2}
\begin{equation}
\rho(\vec{q},i\omega_{n})=\frac{\delta \tilde{S}}{i\delta A_0} =
K_{00}(\vec{q},i\omega_{n})i\delta A_{0}(\vec{q},i\omega_{n}) \;\;.
\end{equation}
For $i\omega \neq 0$,
\begin{equation}\label{K00}
K_{00}(\vec{q},i\omega_{n})=\frac{\tilde{n}}{2\pi\tilde{\phi}det
D(\vec{q},i\omega_{n})} \frac{q^2}{\tilde{\omega}_c}
(\Sigma_{4}^{2}-\Sigma_{0}\Sigma_{6}) \;\;.
\end{equation}
The linear response is obtained by the standard procedure
$i\omega_{n} \rightarrow \omega + i\eta$. Thus one may find
the collective excitation spectrum by locating the poles in
$K_{00}$. Notice that if one applies a similar procedure
to a system without the constraint, one would obtain the same
response kernel $K_{00}$ as that with an arbitrary
$\delta A_\mu$. Hence it is reasonable to expect that $K_{00}$
gives the correct density-density correlation, and is sufficient
for the purpose of studying physics other than transport
properties.

The compressibility is obtained from
\begin{equation}
\kappa = \lim_{q \rightarrow 0} K_{00}(\vec{q},i\omega_{n}=0)=
\lim_{q \rightarrow 0}
\frac{\tilde{n}}{2\pi\tilde{\phi}det D^{0}(\vec{q})}
\frac{q^2}{\tilde{\omega}_c}
((\Sigma_{4}^0)^{2}-\Sigma_{0}^0\Sigma_{6}^0) \;\;,
\end{equation}
where $D^{0}(\vec{q})$ is obtained from (\ref{detD}) with
the substitution $\Sigma_{j} \rightarrow \Sigma^{0}_{j}$.
At $T=0$, $\kappa =0$, the system is incompressible. As $T$
is raised from zero, the compressibility acquires an
exponential correction as expected.

\section{Discussions and Concluding Remarks}

The mass renormalization equation (\ref{massrn}) can in principle
be solved numerically, which is quite complicated due to the large
(infinite) number of singularities in the integrand. We shall not
discuss it the present work. Instead, we will offer a conjecture
on a solution based on physical considerations. Since the physical
energy scale is given by the electron-electron interaction, we have
\begin{equation}
\frac{eB}{m^{*}c} \propto \frac{e^2}{\epsilon l_0} \;\;,
\end{equation}
when $v(\vec{r})$ is the Coulomb interaction.
Here $l_0$ is the bare magnetic length,
$l_{0} = \sqrt{\frac{\nu}{n}}\tilde{l}_{0}$.
We may in this case write ($\hbar=1$)
\begin{equation}\label{gamma}
\bar{\lambda}=-1+\gamma \frac{e^{2}/\epsilon l_0}{\omega_c} \;\;,
\end{equation}
where $\gamma$ is a numerical factor (to be determined from
(\ref{massrn})).

A strict constraint on any theory of electrons in magnetic field
is the celebrated Kohn's theorem, which states that in an impurity
free system the total oscillation strength $K_{00}$ is
saturated by the cyclotron mode (to the order $q^2$), regardless
the details of interactions. Since in our theory the system
is constrained to the LLL, the cyclotron mode is pushed up
to infinite. Kohn's theorem then requires that all the modes
in our theory must be weighted with some higher power then $q^2$
as $q \rightarrow 0$. To ensure the one-loop expansion adopted
here to be sensible, we must demonstrate that Kohn's theorem is
indeed satisfied within our Gaussian theory. Below we show
this explicitly for the case $T=0$.

As $\beta \rightarrow \infty$, we have
\begin{eqnarray}
\Sigma_{j} & = & \frac{e^{-x}}{n}\sum_{m=0}^{n-1}\sum_{l=n}^{\infty}
\frac{l-m}{(\frac{\omega}{\tilde{\omega}_{c}})^{2}-(l-m)^{2}}
\frac{m!}{l!}x^{l-m-1}
\left [ L_{m}^{l-m}(x) \right ]^{2-j} \left [
(l-m-x)L_{m}^{l-m}(x)+2x\frac{dL_{m}^{l-m}(x)}{dx} \right ] ^{j}
\end{eqnarray}
for j=0,1,2; and
\begin{eqnarray}
\Sigma_{3} & = & \frac{e^{-x}}{n}\sum_{m=0}^{n-1}\sum_{l=n}^{\infty}
\frac{l-m}{(\frac{\omega}{\tilde{\omega}_{c}})^{2}-(l-m)^{2}}
\frac{m!}{l!}x^{l-m} \nonumber \\
& & \left [ 2\frac{dL_{m}^{l-m}(x)}{dx} - L_{m}^{l-m}(x) \right ]
\left [ (2(l-m)-x)L_{m}^{l-m}(x)+2x\frac{dL_{m}^{l-m}(x)}{dx} \right ]
\;\;,
\end{eqnarray}
\begin{eqnarray}
\Sigma_{4}=\frac{e^{-x}}{n}
\sum_{m=0}^{n-1}\sum_{l=n}^{\infty} \frac{l-m}
{(\frac{\omega}{\tilde{\omega}_{c}})^{2}-(l-m)^{2}} \frac{m!}{l!}
x^{l-m-1} \left (1-\frac{l+m+1}{1+\tilde{n}} \right )
\left [ L_{m}^{l-m}(x) \right ]^{2} \;,
\end{eqnarray}
\begin{eqnarray}
\Sigma_{5} & = & \frac{e^{-x}}{n}\sum_{m=0}^{n-1}\sum_{l=n}^{\infty}
\frac{l-m}{(\frac{\omega}{\tilde{\omega}_{c}})^{2}-(l-m)^{2}}
\frac{m!}{l!}x^{l-m-1} \nonumber \\
& & \left ( 1-\frac{l+m+1}{1+\tilde{n}} \right ) L_{m}^{l-m}(x)
\left [ (l-m-x)L_{m}^{l-m}(x)+2x\frac{dL_{m}^{l-m}(x)}{dx} \right ]
\;\;,
\end{eqnarray}
\begin{eqnarray}
\Sigma_{6}=\frac{e^{-x}}{n}\sum_{m=0}^{n-1}\sum_{l=n}^{\infty}
\frac{l-m}{(\frac{\omega}{\tilde{\omega}_{c}})^{2}-(l-m)^{2}}
\frac{m!}{l!}x^{l-m-1}
\left ( 1-\frac{l+m+1}{1+\tilde{n}} \right )^2
\left [ L_{m}^{l-m}(x) \right ]^{2} \;,
\end{eqnarray}
For $j=0,1,2,3$, they are just the results obtained previously
for the unconstrained theory at $T=0$ \cite{LF}. As
$q \rightarrow 0$, $\Sigma_j$ may be expanded
to the order $x=\tilde{l}_{0}^{2}q^{2}/2$:
\begin{equation}
\Sigma_{0}=\frac{1}{(\frac{\omega}{\tilde{\omega}_{c}})^{2}-1}
+ \frac{3n}{((\frac{\omega}{\tilde{\omega}_{c}})^{2}-1)
((\frac{\omega}{\tilde{\omega}_{c}})^{2}-4)} x + ... \;\;,
\end{equation}
\begin{equation}
\Sigma_{1}=\frac{1}{(\frac{\omega}{\tilde{\omega}_{c}})^{2}-1}
+ \frac{6n}{((\frac{\omega}{\tilde{\omega}_{c}})^{2}-1)
((\frac{\omega}{\tilde{\omega}_{c}})^{2}-4)} x + ... \;\;,
\end{equation}
\begin{equation}
\Sigma_{2}=\frac{1}{(\frac{\omega}{\tilde{\omega}_{c}})^{2}-1}
+ \frac{n((\frac{\omega}{\tilde{\omega}_{c}})^{2}+8)}
{((\frac{\omega}{\tilde{\omega}_{c}})^{2}-1)
((\frac{\omega}{\tilde{\omega}_{c}})^{2}-4)} x + ... \;\;,
\end{equation}
\begin{equation}
\Sigma_{3}=-\frac{2n}{(\frac{\omega}{\tilde{\omega}_{c}})^{2}-1} x
+ ... \;\;,
\end{equation}
\begin{equation}
\Sigma_{4}=\frac{1}{(\frac{\omega}{\tilde{\omega}_{c}})^{2}-1}
\frac{1+{\tilde n}-2n}{1+{\tilde n}} +
\frac{1-(\frac{\omega}{\tilde{\omega}_{c}})^{2}
+3n(1+{\tilde n}-2n)}{((\frac{\omega}{\tilde{\omega}_{c}})^{2}-1)
((\frac{\omega}{\tilde{\omega}_{c}})^{2}-4)(1+{\tilde n})} x + ...
\;\;,
\end{equation}
\begin{equation}
\Sigma_{5}=\frac{1}{(\frac{\omega}{\tilde{\omega}_{c}})^{2}-1}
\frac{1+{\tilde n}-2n}{1+{\tilde n}} +
2 \frac{1-(\frac{\omega}{\tilde{\omega}_{c}})^{2}
+3n(1+{\tilde n}-2n)}{((\frac{\omega}{\tilde{\omega}_{c}})^{2}-1)
((\frac{\omega}{\tilde{\omega}_{c}})^{2}-4)(1+{\tilde n})} x + ...
\;\;,
\end{equation}
\begin{equation}
\Sigma_{6}=\frac{1}{(\frac{\omega}{\tilde{\omega}_{c}})^{2}-1}
\left ( \frac{1+{\tilde n}-2n}{1+{\tilde n}} \right ) ^2 +
\frac{(5n-2{\tilde n}-2)((\frac{\omega}{\tilde{\omega}_{c}})^{2}-1)
+ 3n(1+{\tilde n}-2n)^{2} }{((\frac{\omega}{\tilde{\omega}_{c}})^{2}-1)
((\frac{\omega}{\tilde{\omega}_{c}})^{2}-4)(1+{\tilde n})^{2}} x + ...
\;\;.
\end{equation}
Substitute these results to (\ref{K00}), we obtain the small $q$
behavior of the density-density response kernel
\begin{equation}\label{K00q0}
K_{00}(q,\omega) = \frac{\tilde{n}}{4\pi\tilde{\phi}}
\frac{q^{4}\tilde{l}_{0}^{2}}{\tilde{\omega}_c}
\frac{\tilde{n}}{(1+\tilde{n}-2n)^{2}}
\frac{1}{(\frac{\omega}{\tilde{\omega}_{c}})^{2}-4} + ...  \;\;.
\end{equation}
Thus one finds that the weight of the poles in $K_{00}$
vanishes according to $q^4$, satisfying Kohn's theorem.
This feature of our theory is non-trivial and is quite
encouraging, indicates strongly that this formalism may have
well captured the correct low energy physics of the FQHE. Notice
that the situation here is quite different from that in the
unconstrained theory \cite{LF} in which the cyclotron mode is
explicitly present and sets the energy scale of all the low
energy excitations. Incidentally, (\ref{K00q0}) also shows
that the energy gap at $q=0$ (at $T=0$) is $2\tilde{\omega}_c
= (\frac{2\gamma}{1+\tilde{n}})\frac{e^2}{\epsilon l_0} $.
{}From our knowledge about the collective energy gap obtained
through numerical works \cite{book,HSH,GMP}, we may
estimate that $\gamma \approx 0.3$ for $\nu = 1/3$.

One may also calculate the dispersion of the low energy
collective modes at finite $q$ by finding the poles of $K_{00}$.
It depends on the details (such as the range and strength) of
the electron-electron interaction, as well as $\bar\lambda$.
The numerical calculation of the full dispersion of the
collective modes is straightforward once the value of
$\bar\lambda$ is known.
However, it is not clear that the above estimate
for $\gamma$ is correct within the {\em one-loop}
expansion. It is well possible (especially in view of the
dynamical nature of the $\lambda$-field) that one needs
corrections beyond the one-loop level to give a reasonable
value for $\gamma$. These issues, of course, can only be clarified
by actually solving the stationary point equation (\ref{massrn}),
which we hope to be done in the near future. In the absence of
the solution for the mass renormalization equation (\ref{massrn}),
we shall not address the quantitative questions such as the
dispersion of the collective modes in the present work.

To summarize, we present in this paper a fermion Chern-Simons
field theory describing two-dimensional electrons in the LLL.
The constraint is realized through a $\delta$-functional
represented by an auxiliary field $\lambda$ which fixes the
kinetic energy of electrons at $\frac{1}{2}\hbar\omega_{c}$.
In a mean-field approximation, the ground state is highly
degenerate for non-interacting systems when $\nu$ is not an
integer. Assuming a new non-degenerate mean field ground
state at $\nu = n/(2pn+1)$ (which is described by $n$ filled
effective LLs) for interacting electrons,
we perform the one-loop expansion around this mean-field
state and subject our assumption to
an self-consistency requirement. The resulting dressed
stationary point equation for $\bar\lambda$ is then physically
interpreted as the mass renormalization equation for the
composite fermions. An important issue which needs to be
investigated in the future is to numerically solve the
renormalization equation to provide an answer for the
the energy gap of a FQH state in the thermodynamical limit.

\begin{center}
{\bf Acknowledgement}
\end{center}

The author thanks G.S. Canright, M. Ma and especially J.K. Jain
for many stimulating discussions, and W. Chen, S.M. Girvin,
A.H. MacDonald, J. Quinn, Z.B. Su and Y.S. Wu for helpful comments.
This work was supported by the National Science Foundation
under Grant No. DMR-9413057, and by the U.S. Department of Energy
through Contract No. DE-AC05-84OR21400 administered by Martin
Marietta Energy Systems Inc..

\appendix{Appendix}

In this appendix, we give some details of calculations of the matrix
elements in $\Pi^{0}$. At zero $T$ and without the constraint, this
has been done in \cite{LF}. Calculations at finite $T$ is quite
similar \cite{LZ,RSS}. The imaginary time free fermion propagator,
written in the Landau gauge, is given by
\begin{equation}
G(r_{1}, r_{2})=\frac{1}{\beta}\sum_{i\omega_{n}}\sum_{l}\sum_{k}
e^{-i\omega_{n}(\tau_{1}-\tau_{2})}
\frac{1}{L\tilde{l}_{0}2^{l}\sqrt{\pi}l!}
e^{ik(x_{1}-x_{2})}\frac{1}{i\omega_{n}-(\epsilon_{l}-\tilde\mu)}
\tilde{H}_{l}(\frac{y_{1}}{\tilde{l}_{0}}-\tilde{l}_{0}k)
\tilde{H}_{l}(\frac{y_{2}}{\tilde{l}_{0}}-\tilde{l}_{0}k)
\;\;,
\end{equation}
with
\begin{equation}
\tilde{H}_{l}(y)=e^{-\frac{y^{2}}{2}}H_{l}(y) \;\;,
\end{equation}
where $H_{l}(y)$ is a Hermit polynomial.
The one-loop kernel in real space may be written as
\begin{equation}
\tilde{\Pi}^{0}(r_{1},r_{2})=-\left(
\begin{array}{cccc}
\langle \rho(r_{1})\rho(r_{2})\rangle &
\langle \rho(r_{1})j_{x}(r_{2})\rangle &
\langle \rho(r_{1})j_{y}(r_{2})\rangle &
\frac{\omega_{c}}{2}\langle \rho(r_{1})\rho(r_{2})\rangle \\
 & & & -\frac{m^*}{m_b}\langle \rho(r_{1})h(r_{2})\rangle  \\
\langle j_{x}(r_{1})\rho(r_{2})\rangle &
\langle j_{x}(r_{1})j_{x}(r_{2})\rangle + \frac{\bar{\rho}}{m^*} &
\langle j_{x}(r_{1})j_{y}(r_{2})\rangle &
\frac{\omega_{c}}{2}\langle j_{x}(r_{1})\rho(r_{2})\rangle  \\
 & & & -\frac{m^*}{m_b}\langle j_{x}(r_{1})h(r_{2})\rangle  \\
\langle j_{y}(r_{1})\rho(r_{2})\rangle &
\langle j_{y}(r_{1})j_{x}(r_{2})\rangle &
\langle j_{y}(r_{1})j_{y}(r_{2})\rangle + \frac{\bar{\rho}}{m^*} &
\frac{\omega_{c}}{2}\langle j_{y}(r_{1})\rho(r_{2})\rangle  \\
 & & & -\frac{m^*}{m_b}\langle j_{y}(r_{1})h(r_{2})\rangle  \\
\frac{\omega_{c}}{2}\langle \rho(r_{1})\rho(r_{2})\rangle &
\frac{\omega_{c}}{2}\langle \rho(r_{1})j_{x}(r_{2})\rangle &
\frac{\omega_{c}}{2}\langle \rho(r_{1})j_{y}(r_{2})\rangle &
\langle (\frac{\omega_{c}}{2}\rho(r_{1})-\frac{m^*}{m_b} h(r_{1})) \\
-\frac{m^*}{m_b}\langle h(r_{1})\rho(r_{2}))\rangle &
-\frac{m^*}{m_b}\langle h(r_{1})j_{x}(r_{2})\rangle &
-\frac{m^*}{m_b}\langle h(r_{1})j_{y}(r_{2})\rangle &
(\frac{\omega_{c}}{2}\rho(r_{2})-\frac{m^*}{m_b} h(r_{2})) \rangle
\end{array}\right) \;\;\;,
\end{equation}
where $ \langle \; \rangle $ denotes time-ordered average.
To make our notation compact, we express the matrix elements of
$\tilde{\Pi}^0$ with index $\mu\nu$, such that for $\mu,\nu = 0,1,2$,
they represent the fluctuating gauge field $a_\mu$, while for
$\mu,\nu = 3$, they represent the fluctuating $\lambda$-field.
Now, as an example, we work out $\tilde{\Pi}_{11}$ in detail.
\begin{eqnarray}
\tilde{\Pi}_{11}(r_{1},r_{2}) & = &
-\frac{1}{4m_{b}^{2}} \{
[D_{x_1}G(r_{1}-r_{2})]D_{x_2}G(r_{2}-r_{1})+
[D_{x_2}^{\dagger}G(r_{1}-r_{2})]D_{x_1}^{\dagger}G(r_{2}-r_{1})
 \nonumber \\
& & -[D_{x_1}D_{x_2}^{\dagger}G(r_{1}-r_{2})]G(r_{2}-r_{1})-
G(r_{1}-r_{2})D_{x_1}^{\dagger}D_{x_2}G(r_{2}-r_{1}) \}
\nonumber \\
& & + \delta^{3}(r_{1}-r_{2})\frac{\overline{\rho}}{m^{*}}
\;\;,
\end{eqnarray}
where $\vec{D}$ is the covariant derivative. In the Landau gauge
it has the form
\begin{equation}
\vec{D} = \vec{\nabla} + i\frac{e}{c}\vec{A} =
\hat{x}(\partial_{x}-i\frac{Be}{c}y)+\hat{y}\partial_{y} \;\;.
\end{equation}
Define the Fourier transform
\begin{equation}
\tilde{\Pi}^{0}_{\mu\nu}(\vec{q},i\omega_{n};\vec{q}\;',i\omega_{n}')=
\int_{0}^{\beta}d\tau_{1}\int_{0}^{\beta}d\tau_{2}
\int d^{2}r_{1}d^{2}r_{2} e^{-i(\vec{q}\vec{r}_{1}-\omega_{n}\tau_{1})}
e^{-i(\vec{q}\;'\vec{r}_{2}-\omega_{n}'\tau_{2})}
\tilde{\Pi}^{0}_{\mu\nu}(r_{1},r_{2}) \;\;.
\end{equation}
With the above expression of the free fermion propagator, we have
\begin{eqnarray}\label{Pi11}
\lefteqn{
\tilde{\Pi}_{11} (\vec{q},i\omega_{n};\vec{q}\;',i\omega_{n}') =
\delta(q_{x}+q_{x}')\delta(i\omega_{n}+i\omega_{n}')
} \nonumber \\
& \left \{
(-\frac{1}{4m_{b}^{2}})\sum_{l,l'}\sum_{k}\sum_{i\nu_n}
\frac{1}{\tilde{l}_{0}^{2}\pi 2^{l}l!2^{l'}l'!}
\frac{1}{i\nu_{n}-(\epsilon_{l}-\tilde\mu)}
\frac{1}{(i\nu_{n}+i\omega_{n})-(\epsilon_{l'}-\tilde\mu)}
I_{1x}(\vec{q})I_{1x}(\vec{q}\;') +
\delta(q_{y}+q_{y}')\frac{\overline{\rho}}{m^{*}}
\right \} \;\;,
\end{eqnarray}
with
\begin{eqnarray}\label{I1x}
I_{1x}(\vec{q}) & = & \int dy e^{-iq_{y}y}
\{ (i(k-q_{x})-\frac{i}{\tilde{l}_{0}^{2}}y)
\tilde{H}_{l}(\frac{y}{\tilde{l}_{0}}-\tilde{l}_{0}k)
\tilde{H}_{l'}(\frac{y}{\tilde{l}_{0}}-\tilde{l}_{0}(k-q_{x})) -
\nonumber \\
& & (-ik+\frac{i}{\tilde{l}_{0}^{2}}y)
\tilde{H}_{l}(\frac{y}{\tilde{l}_{0}}-\tilde{l}_{0}k)
\tilde{H}_{l'}(\frac{y}{\tilde{l}_{0}}-\tilde{l}_{0}(k-q_{x})) \}
\;\;.
\end{eqnarray}
In the above expressions $i\omega_{n}=i\frac{2n\pi}{\beta}$ is the
bosonic Matsubara frequency while $i\nu_n$ is fermionic, given by
$i\frac{(2n+1)\pi}{\beta}$. The sum over $i\nu_n$  may be done
through the standard contour integral method \cite{Mahan}.
The result is (for $i\omega_{n} \neq 0$)
\begin{equation}\label{Msum1}
\sum_{i\nu_n} \frac{1}{i\nu_{n}-(\epsilon_{l}-\tilde\mu)}
\frac{1}{(i\nu_{n}+i\omega_{n})-(\epsilon_{l'}-\tilde\mu)}
=\beta \frac{f(\epsilon_{l'}-\tilde\mu)-f(\epsilon_{l}-\tilde\mu)}
{i\omega_{n}-(\epsilon_{l}-\epsilon_{l'})} \;\;,
\end{equation}
where $f(z)=1/(e^{\beta z}+1)$ is the fermi distribution function.
Now we compute the integral $I_{1x}(\vec{q})$. Using the relation
\cite{table}
\begin{equation}\label{hermit1}
yH_{l}(y)=\frac{1}{2}H_{l+1}(y)+lH_{l-1}(y) \;\;,
\end{equation}
we have
\begin{eqnarray}
I_{1x}(\vec{q}) & = & \int dy e^{-iq_{y}y} (-\frac{i}{\tilde{l}_{0}})
\{ \tilde{H}_{l}(\frac{y}{\tilde{l}_{0}}-\tilde{l}_{0}k)
\left [ \frac{1}{2} \tilde{H}_{l'+1}(\frac{y}{\tilde{l}_{0}}-
\tilde{l}_{0}(k-q_{x})) + l'\tilde{H}_{l'-1}(\frac{y}{\tilde{l}_{0}}-
\tilde{l}_{0}(k-q_{x})) \right ] \nonumber \\
& & + \left [ \frac{1}{2} \tilde{H}_{l+1}(\frac{y}{\tilde{l}_{0}}-
\tilde{l}_{0}k) + l\tilde{H}_{l-1}(\frac{y}{\tilde{l}_{0}}-
\tilde{l}_{0}k)  \right ]
\tilde{H}_{l'}(\frac{y}{\tilde{l}_{0}}-\tilde{l}_{0}(k-q_{x})) \}
\;\;.
\end{eqnarray}
Shifting the variable $y$, and use \cite{table}
\begin{equation}\label{hermit2}
\int e^{-x^2}H_{m}(x+y)H_{l}(x+z)dx=\sqrt{\pi}2^{l}m!z^{l-m}
L^{l-m}_{m}(-2yz)\;\;,\;\;\;\; m \leq l \;\;,
\end{equation}
we have
\begin{eqnarray}\label{I1x1}
I_{1x}(\vec{q}) & = & (-\frac{i}{\tilde{l}_{0}})
e^{-iq_{y}\tilde{l}_{0}^{2}(k-\frac{q_{x}}{2})}
e^{-\frac{\tilde{l}_{0}^{2}q^{2}}{4}} \pi\tilde{l}_{0}2^{l'}l!
(\frac{\tilde{l}_{0}(q_{x}-iq_{y})}{2})^{l'-l-1}
\nonumber \\
& & \left \{ \frac{\tilde{l}_{0}^{2}(q_{x}-iq_{y})^{2}}{4}
L^{l'-l+1}_{l}(x) +\frac{l'}{2}L^{l'-l-1}_{l}(x)+\frac{1}{2}(l+1)
L^{l'-l-1}_{l+1}(x) + \frac{\tilde{l}_{0}^{2}(q_{x}-iq_{y})^{2}}{4})
L^{l'-l+1}_{l-1}(x) \right \}  \;\;,
\nonumber \\ & & (l<l') \;\;;
\nonumber \\
& = & \left( \; l \leftrightarrow l' \;\;,\;\;\;
q_{x}-iq_{y} \leftrightarrow -(q_{x}+iq_{y}) \;\; \right ) \;\;,
\nonumber  \\ & & (l>l') \;\;.
\end{eqnarray}
Here $x \equiv \tilde{l}_{0}^{2}q^{2}/2$.
After some algebra, we obtain
\begin{eqnarray}
\frac{\tilde{l}_{0}(q_{x}+iq_{y})}{2}I_{1x}(\vec{q}) & = &
(-\frac{i}{\tilde{l}_{0}})
 e^{-iq_{y}\tilde{l}_{0}^{2}(k-\frac{q_{x}}{2})}
e^{-\frac{\tilde{l}_{0}^{2}q^{2}}{4}} \pi\tilde{l}_{0}2^{l'}l!
(\frac{\tilde{l}_{0}(q_{x}-iq_{y})}{2})^{l'-l-1}
 \nonumber \\
& & \left \{ q_{x}(l'-l)L^{l'-l}_{l}(x)+iq_{y} \left [
2x\frac{dL^{l'-l}_{l}(x)}{dx} + (l'-l-x)L^{l'-l}_{l}(x) \right ]
\right \} \;, \;\;\;  (l<l')\;;
\nonumber \\
-\frac{\tilde{l}_{0}(q_{x}-iq_{y})}{2}I_{1x}(\vec{q}) & = &
(\frac{i}{\tilde{l}_{0}})
 e^{-iq_{y}\tilde{l}_{0}^{2}(k-\frac{q_{x}}{2})}
e^{-\frac{\tilde{l}_{0}^{2}q^{2}}{4}} \pi\tilde{l}_{0}2^{l}l'!
(-\frac{\tilde{l}_{0}(q_{x}+iq_{y})}{2})^{l-l'-1}
 \nonumber \\
& & \left \{ q_{x}(l-l')L^{l-l'}_{l'}(x)-iq_{y} \left [
2x\frac{dL^{l-l'}_{l'}(x)}{dx} + (l-l'-x)L^{l-l'}_{l'}(x) \right ]
\right \} \;, \;\;\;  (l>l')\;;
\end{eqnarray}
The summation over $k$ may be straightforwardly carried out
\begin{equation}
\sum_{k}e^{i\tilde{l}_{0}^{2}k(q_{y}+q_{y'})}=
\frac{L^{2}}{2\pi\tilde{l}_{0}^{2}}\delta (q_{y}+q_{y}') \;\;,
\end{equation}
where $\frac{L^{2}}{2\pi\tilde{l}_{0}^{2}}$ is the degeneracy of a
(pseudo) Landau level. These lead to
\begin{eqnarray}
\lefteqn{
\tilde{\Pi}^{0}_{11}(\vec{q},i\omega_{n};\vec{q}\;',i\omega_{n}')=
\delta^{2}(\vec{q}+\vec{q}\;')\delta(i\omega_{n}+i\omega_{n}')  (
} \nonumber \\
& \frac{L^{2}\beta}{2\pi\tilde{\omega}_{c}}e^{-x}\sum_{m<l}
\frac{l-m}{(\frac{i\omega_n}{\tilde{\omega}_{c}})^{2}-(l-m)^{2}}
\frac{m!}{l!}x^{l-m-1}
\{ f(\epsilon_{m}-\tilde\mu)-f(\epsilon_{l}-\tilde\mu) \} \nonumber \\
& \left \{
(i\omega_{n})^{2}(L^{l-m}_{m}(x))^{2} +
x \left [ 2\frac{dL_{m}^{l-m}(x)}{dx} - L_{m}^{l-m}(x) \right ]
\left [ (2(l-m)-x)L_{m}^{l-m}(x)+2x\frac{dL_{m}^{l-m}(x)}{dx} \right ]
\tilde{\omega}_{c}^{2} \frac{q_{y}^2}{q^2}
\right \}
\nonumber \\
& -\frac{L^{2}\beta}{2\pi\tilde{\omega}_{c}}e^{-x}\sum_{m<l}
\frac{m!}{l!}x^{l-m-1}(l-m)\{
f(\epsilon_{m}-\tilde\mu)-f(\epsilon_{l}-\tilde\mu) \}
\tilde{\omega}_{c}^{2}(L^{l-m}_{m}(x))^{2} +
\frac{\overline{\rho}}{m^{*}}  ) \;\;.
\end{eqnarray}
Define
\begin{equation}
\tilde{\Pi}^{0}(\vec{q},i\omega_{n};\vec{q}\;',i\omega_{n}')
\equiv \delta(q_{x}+q_{x}')\delta(i\omega_{n}+i\omega_{n}')
\Pi^{0}(\vec{q},i\omega_{n}) \;\;,
\end{equation}
we have
\begin{eqnarray}
\Pi^{0}_{11}(\vec{q},i\omega_{n}) & = &
\frac{\tilde{n}}{2\pi \tilde{\phi}} \left \{
\frac{i\omega_{n}^2}{\tilde{\omega}_c}\Sigma_{0}
+\tilde{\omega}_{c} \frac{q_{y}^{2}}{q^2}\Sigma_{3} \right \} -
\nonumber \\
& & \frac{\tilde n}{2\pi\tilde{\phi}} \tilde{\omega}_{c}
\frac{e^{-x}}{n}\sum_{m<l} \frac{m!}{l!}x^{l-m-1}(l-m)
[ f(\epsilon_{m}-\tilde\mu)-f(\epsilon_{l}-\tilde\mu) ] \;\;.
\end{eqnarray}
The last term vanishes as $x \rightarrow 0$. According to \cite{LF},
we ignore it hereafter. (A similar treatment is also applied to
$\Pi^{0}_{22}$.) This ensures the gauge invariance of $\Pi^0$.
The new matrix elements which appear in the present theory are
those involve the (mean-field) kinetic energy density $h(r)$. Now we
evaluate one of them as an example.  Consider
\begin{eqnarray}
\tilde{\Pi}_{03}(r_{1},r_{2}) & = & \frac{\omega_{c}^b}{2}
G(r_{1}-r_{2})G(r_{2}-r_{1})+
\nonumber \\
& & \frac{1}{4m_b} \left \{ G(r_{1}-r_{2})
{\vec D}_{\vec{r}_2}^{2}G(r_{2}-r_{1}) +
[ {\vec D}_{\vec{r}_2}^{\dagger 2}G(r_{1}-r_{2}) ]G(r_{2}-r_{1})
\right \} \;\;.
\end{eqnarray}
In the Fourier space we have
\begin{eqnarray}
\tilde{\Pi}_{03}(\vec{q},i\omega_{n};\vec{q}\;',i\omega_{n}')
& = & \delta(q_{x}+q_{x}')\delta(i\omega_{n}+i\omega_{n}')
\sum_{l,l'}\sum_{k}\sum_{i\nu_n}
\frac{1}{\tilde{l}_{0}^{2}\pi 2^{l}l!2^{l'}l'!}
\nonumber \\
& & \frac{1}{i\nu_{n}-(\epsilon_{l}-\tilde\mu)}
\frac{1}{(i\nu_{n}+i\omega_{n})-(\epsilon_{l'}-\tilde\mu)} \left (
\frac{\omega_{c}^b}{2}  -\frac{l+l'+1}{2m_{b}\tilde{l}_{0}^2}
\right ) I_{0}(\vec{q})I_{0}(\vec{q}\;') \;\;,
\end{eqnarray}
with
\begin{eqnarray}
I_{0}(\vec{q}) = \int dy e^{-iq_{y}y} \tilde{H}_{l}
(\frac{y}{\tilde{l}_{0}}-\tilde{l}_{0}k) \tilde{H}_{l'}
(\frac{y}{\tilde{l}_{0}}-\tilde{l}_{0}(k-q_{x})) \;\;.
\end{eqnarray}
$I_0$ may be evaluated straightforwardly using (\ref{hermit2}).
The final result is given in (\ref{Pi0})

Now we consider the case $i\omega_{n} = 0$. For $l' \neq l$, the
Matsubara frequency sum in (\ref{Pi11}) is still given by
(\ref{Msum1}) with substitution $i\omega_{n} = 0$. However, the
term with $l' = l$ also contribute in the present situation:
\begin{equation}\label{Msum2}
\sum_{i\nu_n} \frac{1}{[i\nu_{n}-(\epsilon_{l}-\tilde\mu)]^2}
=\beta \frac{\partial f(z)}{\partial z}|_{z=\epsilon_{l}-\tilde\mu}
=-\beta^{2} e^{(\epsilon_{l}-\tilde\mu)}f^{2}(\epsilon_{l}-\tilde\mu)
\;\;.
\end{equation}
To evaluate, e.g., $\Pi^{0}_{11}(q,0)$, one needs calculate the
integral (\ref{I1x}) for the case $l=l'$. Using a similar procedure,
one obtains
\begin{equation}
I_{1x}(\vec{q})|_{l=l'}=
q_{y}e^{-iq_{y}\tilde{l}_{0}^{2}(k-\frac{q_{x}}{2})}
e^{-\frac{\tilde{l}_{0}^{2}q^{2}}{4}} \pi\tilde{l}_{0}2^{l}l!
\left [ L^{0}_{l}(x)-2L^{1}_{l}(x) \right ] \;\;.
\end{equation}
This result, together with that obtained for $l \neq l'$, eqn.
(\ref{I1x1}), gives $\Pi^{0}_{11}(q,0)$ expressed in (\ref{Pi00}).
Other matrix elements in (\ref{Pi0}) and (\ref{Pi00}) may be
obtained in similar manner.

\figure{
Some mean field ground state configurations at $\nu=1/3$ in the
theory with the LLL constraint. $E$ is in the unit of
$\hbar \bar{\omega}_c$. In the configuration (a), all the
fermions are filled into the first LL to satisfy (\ref{meanE}).
However, there are other ways of placing fermions which also
produce homogeneous states satisfying (\ref{meanE}), such as
those shown in (b),(c) and (d), in which some fermions are
placed in the 0-th LL and higher LLs. The degeneracy of the
mean field ground state becomes infinite in the thermal
dynamical limit.
}


\begin{references}
\bibitem{fnote2}
Although $\Sigma_2$ does not appear in $\Pi^0$, we still show it here
and in the subsequent calculations. $\Sigma_2$ emerges when one adopts
the special guage choice ${\vec \nabla}\cdot \vec{a} =0$, with
$\vec{a}_{T} \parallel \hat{y}$ and $\vec{q} \parallel \hat{x}$, as it
was done in \cite{SH,HSH,LZ} for the unconstrained case. While the
two calculations coincide there in the $q \rightarrow 0$ limit, we find
that they differ at finite $q$, contrary to what was found in \cite{SH}.
The physical origin of this discrepancy is not clear, although one
may argue that it is due to the ``weak form" \cite{LF} of the gauge
invariance property of $\Pi^0$. They would give the same collective
modes (i.e., the same poles in $K_{00}$) if one substitutes
$(\frac{i\omega_n}{\tilde{\omega}_c})^2\Sigma_{0}+\Sigma_{3}$ with
$\Sigma_{2} + 1$. One may also make such a substitution in (\ref{detD})
for the constraint theory. Notice that such a problem does not
occur in the $\nu=1/2$ case considered in \cite{HLR} where the
mean-field ground state is a filled Fermi-sea.
\end{references}
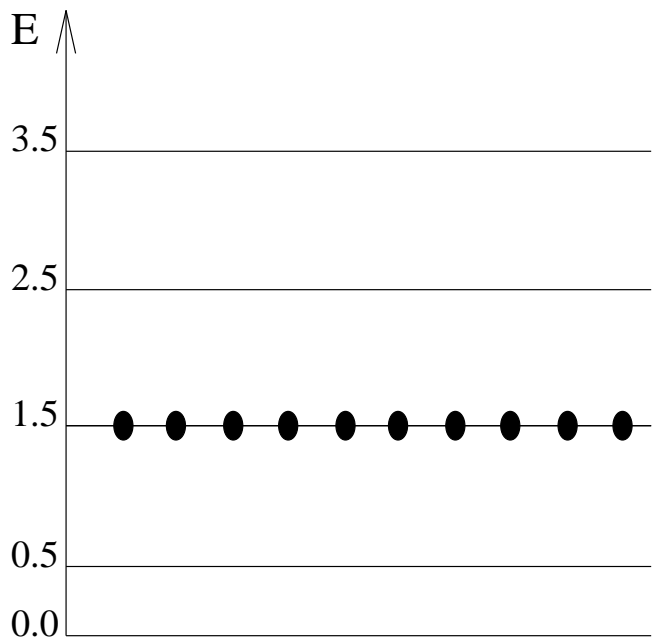
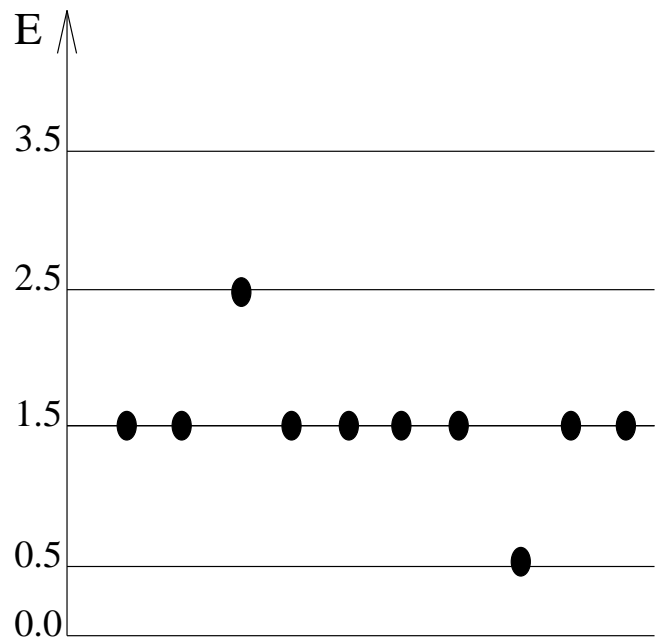
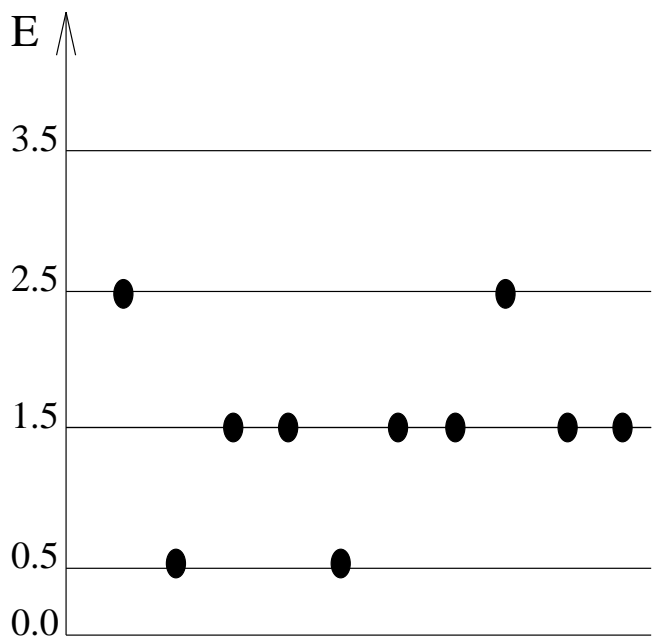
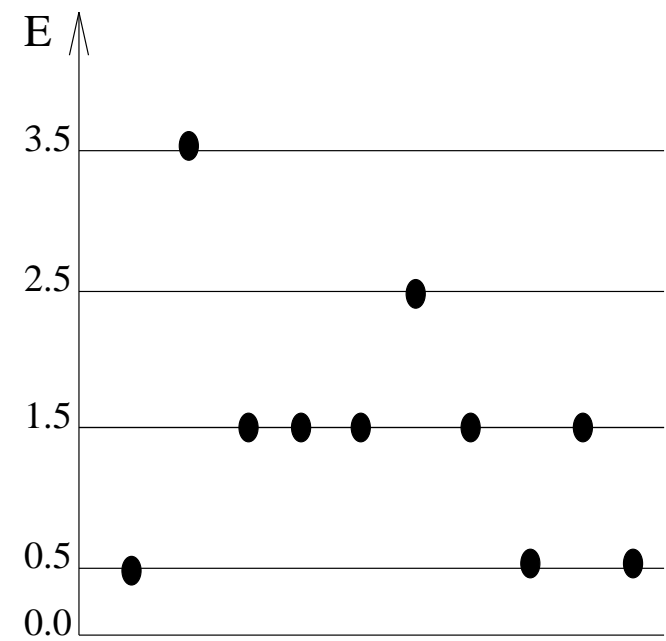
\end{document}